\documentclass{aa}
\usepackage{natbib}
\bibpunct{(}{)}{;}{a}{}{,}

\usepackage{epsfig}
\usepackage{graphics}
\usepackage{float}
\usepackage{amsmath}
\usepackage{multirow}
\usepackage{longtable}
\usepackage{rotate}
\usepackage{amssymb}
\usepackage{array}
\usepackage{times}
\def\phn{\phantom{0}}  
\def\phs{\phantom{$-$}}    

\def\tablecomments#1{\par\smallskip\noindent Notes. #1}
\def\plotone#1{\centerline{\psfig{figure=#1,width=\hsize,clip=}}}
\def\kms{\ifmmode{\rm km\,s^{-1}}\else\hbox{$\rm km\,s^{-1}$}\fi}
\def\nodata{\phs$\cdots$}

\setlongtables

\def\ergs{\,erg\,s$^{-1}$}
\def\phn{\phantom{0}}
\def\phs{\phantom{$-$}}
\def\nodata{\phs$\cdots$}
\def\etal{et~al.}
\def\kms{\,km\,s$^{-1}$}

\def\SII{S\,{\sc ii}}

\def\SiII{Si\,{\sc ii}}
\def\SiIII{Si\,{\sc iii}}
\def\MgII{Mg\,{\sc ii}}
\def\CaII{Ca\,{\sc ii}}
\def\FeII{Fe\,{\sc ii}}

\def\Ang{\,\AA}

\begin{document}

\title{Spectroscopy of the type Ia supernova SN 2002er: days $-$11 to +215 }

\author{R. Kotak\inst{1}
     \and W.P.S. Meikle\inst{1}
     \and G. Pignata\inst{2}
     \and M. Stehle\inst{3}
     \and S.J. Smartt\inst{4}
     \and S. Benetti\inst{5}
     \and W. Hillebrandt\inst{3}
     \and D.J. Lennon\inst{7}
     \and P.A.~Mazzali\inst{3,6}
     \and F.~Patat\inst{2}
     \and M. Turatto\inst{5}
}  

 \offprints{R. Kotak}

\institute{Astrophysics Group, Imperial College London, Blackett Laboratory,
               Prince Consort Road, London, SW7 2AZ, U.K. 
\and European Southern Observatory, Karl-Schwarzschild-Str. 2, D-85748
     Garching bei M\"unchen, Germany
\and Max-Planck-Institut f\"ur Astrophysik, P.O. Box 1317 D-85741 Garching, Germany
\and Department of Physics and Astronomy, Queen's University Belfast, Belfast, BT7 1NN, 
     Northern Ireland, U.K.
\and Osservatorio Astronomico di Padova, vicolo dell'Osservatorio 5, I-35122 Padova, Italy
\and Osservatorio Astronomico di Trieste, Via Tiepolo, 11, I-34131 Trieste, Italy
\and Isaac Newton Group of Telescopes, Apartado de Correos 321, 38700 Santa Cruz de la Palma, 
     Canary Islands, Spain
}

 \date{Received; Accepted; }

\abstract{ We present an extensive set of optical spectroscopy of the
nearby type Ia supernova, SN 2002er, with 24 epochs spanning $-$11 to
+34 days. Its spectral evolution is fairly typical of a type Ia
supernova although it suffers high extinction. Nevertheless, there are
differences in the spectral evolution when compared to coeval spectra
of other normal type Ia supernova with comparable decline-rate
parameters.  Modelling of the photospheric phase spectra using a
homogeneous abundance distribution in the atmosphere provides a fair
match to the observations, but only by pushing the adopted distance
and risetime close to the observational limits. Future improvements
here will require models with a more realtistic stratified abundance
distribution.  From simple modelling of a nebular spectrum obtained at
+215\,d, we infer a $^{56}$Ni mass of 0.69M$_\odot$, consistent with
that derived from the light curve.
}

\authorrunning{Kotak}
   \titlerunning{Spectroscopy of SN 2002er}
   \maketitle

\section{Introduction}

Type Ia supernovae (SNe~Ia) are currently the most accurate high-z
cosmological distance indicator. Through the light curve shape
corrections, accuracies of better than 10\% in distance are claimed
\citep[e.g.][]{phillips:99, riess:98}.
However, a basic assumption is that the correlations hold at all
universal epochs.  Yet, even for the best nearby SN~Ia sample it appears
that the various incarnations of the light curve correction methods
may not be wholly consistent with each other \citep{drell:00,
leibundgut:00}.

Although there is a general consensus that type~Ia supernovae result
from the thermonuclear fusion of C-O white dwarfs, observational
support for the numerous progenitor channels that have been proposed
\citep[e.g.][]{branch:95} is only just beginning to be gathered
\citep[e.g.][]{ruizlapuente:04,hamuy:03,kotak:04}.  In contrast, there
is an increasing body of evidence for the observational diversity
\citep[e.g.][]{benetti:04b} among type Ia supernovae perhaps
suggestive of diversity in the progenitor channels and/or explosion
mechanism.

In the favoured scenario leading to a type~Ia explosion, the white
dwarf is one component of a close binary, and accretes hydrogen from
its companion star. Other scenarios include the merging of binary
white dwarfs, or the accretion of He rather than H \citep[e.g.][]{hn:00}.
When the white dwarf reaches the Chandrasekhar
limit, explosive carbon burning sets in. Burning to nuclear
statistical equilibrium ensues, yielding mostly radioactive
$^{56}$Ni. At lower densities, intermediate-mass nuclei,
e.g. $^{28}$Si, are produced. Thus the typical observed early-time
spectra of SNe~Ia are dominated by lines of both iron-group and
intermediate mass elements. However, the exact nature of the explosion
mechanism is uncertain. Competing models can be tested using detailed
spectral observations since they are highly sensitive to the nature of
the explosion.  For example, much higher velocities in
intermediate-mass elements are observed in delayed-detonation model
spectra than in those of deflagration models (e.g. Khokhlov 1991).
Although the optical region is spectroscopically crowded and
line-identification is difficult due to the additional complication of
blending as a result of Doppler broadening (several thousand
kms$^{-1}$), it is also a region rich in diagnostic potential. For
instance, based on measurements and modelling of lines such as SiII
$\lambda$6355\,{\AA}, CII $\lambda$6578, $\lambda$7231\,{\AA}, and the
CaII H and K lines, \citet{benetti:04a} argue that the combination of
high-velocity intermediate mass elements, an unusually low
temperature, and low abundance of carbon indicate that burning to Si
penetrated to much higher layers than expected for `normal' type~Ia
supernovae. This was interpreted as evidence for a delayed-detonation
explosion. An alternative scenario could be additional nuclear 
burning in the so-called distibuted regime at low densities which
would have a similar effect \citep{roepke:04}.

To date, good photometric and spectroscopic coverage extending from
pre-maximum light to late times has been obtained for only a handful
of supernovae.  In order to better understand and quantify the various
peculiarities that have been observed, high quality photometric and
spectroscopic data from pre-maximum through to nebular epochs is a
prerequisite, with detailed comparisons with the predictions of
explosion models being made at every step.  To this end, the European
Supernova Collaboration (ESC)\footnote{
http://www.mpa-garching.mpg.de/\textasciitilde rtn/} aims to obtain unprecedented
photometric and spectroscopic coverage for $\sim$10 nearby
(v$_{\mathrm{rec}} \la 4000$\,\kms ) type Ia supernovae.  The
potential of high quality early-time spectroscopy is illustrated by
the method of ``abundance tomography" \citep{stehle:04} whereby the
detailed abundance stratification is derived by computing synthetic
spectra for over ten epochs, using spectroscopy from the earliest
epochs to constrain later ones.

We have already obtained excellent data sets for about eight nearby
type~Ia SNe. This is the third in a series of observational papers published on
behalf of the ESC. We present early-time optical spectroscopy spanning
24 epochs of the second ESC target, SN 2002er. In a companion paper,
\citet{gp:04} discuss the photometric evolution of SN 2002er.

\begin{figure}
\plotone{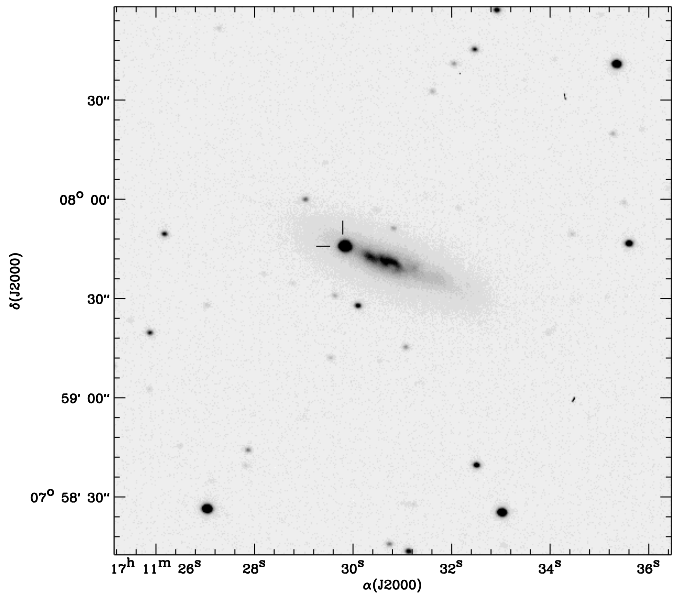}
\caption{$B$ band image of SN 2002er taken around maximum light. The
         dashes indicate the position of the supernova.}
\label{fig:chart}
\end{figure}

\begin{figure*}[!h]
\plotone{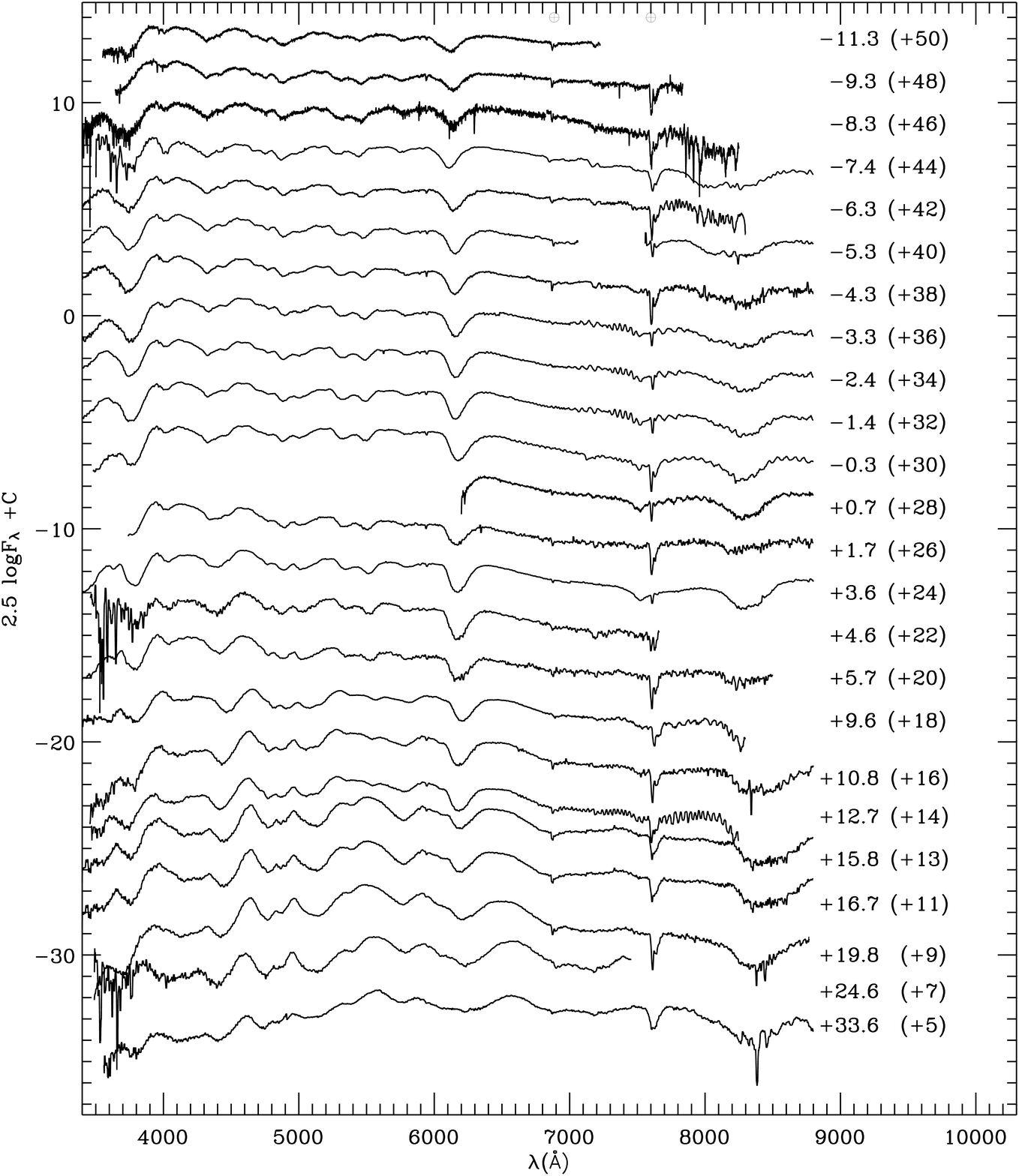}
\caption{Spectral evolution of SN 2002er during the photospheric
phase.  The spectra shown here have neither been corrected for the 
redshift of UGC~ 10743, nor have they been corrected for reddening. The 
spectra have been shifted vertically for clarity; these shifts are indicated
in parenthesis next to the epoch of the spectra (see also Table
\ref{tab:obslog}). 
Note that the quality of the spectra beyond
$\sim$9000\,{\AA} is poor. The prominent telluric features are marked
by earth symbols. }
\label{fig:specevol}
\end{figure*}
\subsection{SN 2002er}

SN~2002er in UGC 10743 was discovered on 2002 Aug. 23.2 U.T. in NEAT
images by \citet{wv:02}. and was spectroscopically confirmed to be a
type~Ia supernova three days later by \citet{ss:02}; comparison with
SN~1994D suggested that it was discovered at an early phase (about
$-10$\,d before maximum light) and that it was significantly redder. A
list of basic parameters is given in Table \ref{tab:par02er}.

\begin{table}
\caption{Properties of SN 2002er and its host galaxy}
\begin{tabular}{ll}
\hline
Host galaxy                        & UGC\,10743      \\
Host galaxy type                   & Sa $^\dag$      \\
RA (2000)                          & $17^h 11^m 29^s.88$  \\
Dec (2000)                         & $+07\degr 59'44''.8$  \\
Recession velocity  (\kms)         & $2569$  \\ 
Distance modulus$^{*}$             & $32.9\pm 0.2$          \\
(H$_0=71$kms$^{-1}$Mpc$^{-1}$)     &                       \\
E$(B-V)^{*}$                       & 0.36                   \\
Offset from nucleus                &       $12''.3$\,W  ~~  $4''.7$\,N       \\
Date of B maximum (MJD)$^{*}$      & 52524.2 (Sep. 07, 2002) \\
Magnitude at maximum$^{*}$         & U=14.72, B=14.89, V=14.59, \\
                                   & R=14.43, I=14.49 \\
$\Delta m_{15}$(B)$^{*}$           & 1.33$\pm$0.04    \\
                                                                
\hline
\end{tabular}
\tablecomments{$^{*}$ Taken from \cite{gp:04}; {\dag} NED.}
\label{tab:par02er}
\end{table}

\section{Data acquisition and reduction}
\label{sec:obs}

The photospheric-phase observations span $-11$ to +34\,days. Somewhat
unusually, the spectroscopic monitoring began several days before the
photometric campaign.  Optical spectra of SN~2002er were obtained from
several different telescopes (Table \ref{tab:obslog}). The
observations were carried out in service or target-of-opportunity mode
usually using the configuration available at the time. This meant that
the set-up varied with epoch and even with observations performed
using the same instrument (see Table \ref{tab:obslog}).

The reduction of the data was carried out using standard NOAO IRAF and
FIGARO routines \citep{shortridge:02}.  The CCD frames were debiased
and corrected for distortion and cosmic rays. Where a sufficient
number of flatfield frames were available, the data were also
flatfielded. Sky subtraction was carried out by fitting low-order
polynomials to the background on either side of the supernova spectrum
and optimally extracting one-dimensional spectra from the
two-dimensional frames. The wavelength solution was derived using arc
frames and was checked against the bright night-sky emission lines.
Flux calibration was carried out with respect to spectrophotmetric
standards usually taken using an identical instrument configuration
and reduced as described above. No attempt was made to correct for the
telluric absorption features between 6000 -- 8000\,{\AA}.

To obtain full wavelength coverage in the optical region, an
observation at any given epoch typically consisted of two or three
spectra taken either with a single grating/grism at different central
wavelengths or using different gratings/grisms.  
Severe distortion between the blue and red parts of the optical 
spectra was apparent on several nights and variable conditions
exacerbated the situation. 
To correct for this distortion we multiplied the fluxed supernova spectra 
by the $BVRI$ Bessell filter functions and integrated across the bands.
A similar procedure was applied to the spectrum of Vega, and hence
spectroscopic magnitudes for the supernova were derived.  Comparison
with near-contemporaneous photometry \citep[from][]{gp:04} was then
used to derive scaling factors for each band.  A low-order polynomial
fit to the scaling factors was obtained which was then applied to
full supernova spectrum.  We estimate the flux calibration to be
accurate to about 10-15\%.  For the earliest epochs ($-$11 to $-$8\,d)
no photometry was available for comparison. Since it is the behaviour
at the earliest epochs that we seek to understand, we did not scale
these spectra in any preconceived manner.

\begin{table*}
\setlength{\tabcolsep}{5pt}
\caption[]{Log of spectroscopic observations of SN 2002er}
 \begin{centering}
 \begin{tabular}{llclllllll}
 \label{tab:obslog}
             &         &             &       &         &  &     &  & \\
\hline
     Date    &  M.J.D. & Epoch$^*$       &  Approx. Range    &  Telescope+Instrument+  &     & Flux   \\
             &         & \,\,\,\,\,(d)   &   \,\,\,\,\,\,\,({\AA})  & Grating/Grism  &   & Standard &     \\
\hline
 20020826    &  52512.93      & \phn$-$11.3         &  3500--7200     & INT+IDS+R400V        &    & HD\,19445     \\
 20020828    &  52514.87      & \phn\phn$-$9.3      &  3600--7800     & INT+IDS+R632V        &    &  Kopff 27     \\
 20020829    &  52515.86      & \phn\phn$-$8.3      &  3300--8900     & INT+IDS+R632V        &    &  \nodata             \\
 20020830    &  52516.84      & \phn\phn$-$7.4      &  3600--9000     & A1.82+AFOSC+gm2,4  &       & BD+28 4211   \\
 20020831    &  52517.89      & \phn\phn$-$6.3      &  3400--9900     & INT+IDS+R632V      &       & Kopff 27     \\
 20020901    &  52518.85      & \phn\phn$-$5.3      &  3200--9100     & CA+CAFOS+B,R200$^\dagger$    &       & BD+28 4211   \\
 20020902    &  52519.88      & \phn\phn$-$4.3      &  3250--9300     & INT+IDS+R632V$^\dagger$      &       & Kopff 27     \\
 20020903    &  52520.85      & \phn\phn$-$3.3      &  3200--10000    & CA+CAFOS+B,R200$^\dagger$    &       & BD+28 4211   \\
 20020904    &  52521.83      & \phn\phn$-$2.4      &  3200--10000    & CA+CAFOS+B,R200$^\dagger$    &       & BD+28 4211   \\
 20020905    &  52522.82      & \phn\phn$-$1.4      &  3200--10000    & CA+CAFOS+B,R200$^\dagger$    &       & BD+28 4211   \\
 20020906    &  52523.85      & \phn\phn$-$0.3      &  3500--10000    & CA+CAFOS+B,R200              &       & BD+28 4211 &       \\
 20020907    &  52524.89      &  \phn\phn+0.7       &  6200--10000    & CA+CAFOS+B,R200$^\ddag$      &       & BD+28 4211  \\
 20020908    &  52525.88      &  \phn\phn+1.7       &  3700--9300     & CA+CAFOS+B,R200$^\dagger$    &       & BD+28 4211   \\
 20020910    &  52527.81      &  \phn\phn+3.6       &  3200--10000    & CA+CAFOS+B,R200$^\dagger$    &       & BD+28 4211   \\
 20020911    &  52528.81      &  \phn\phn+4.6       &  3450--7600     & A1.82+AFOSC+gm4    &         & \nodata         \\
 20020912    &  52529.88      &  \phn\phn+5.7       &  3250--9800     & CA+CAFOS+B,R20$^\dagger$     &       & BD+28 4211   \\
 20020916    &  52533.84      &  \phn\phn+9.6       &  3300--8700     & INT+IDS+R150V      &       & Kopff 27          \\
 20020918    &  52535.03      &  \phn+10.8          &  3450--9600     & D1.54+DFOSC+gm4    &       & LTT 7987          \\
 20020919    &  52536.87      &  \phn+12.7          &  3450--8800     & INT+IDS+R150V$^\dagger$    &       & Kopff 27     \\
 20020922    &  52540.00      &  \phn+15.8          &  3300--9750     & D1.54+DFOSC+gm5    &       & LTT 7987          \\
 20020923    &  52540.86      &  \phn+16.7          &  3300--9700     & INT+IDS+R150V      &       & Kopff 27    \\ 
 20020924    &  52542.00      &  \phn+17.8          &  9400--24900    & NTT+SofI           &       & HIP 95550    \\
 20020926    &  52544.02      &  \phn+19.8          &  3500--8800     & D1.54+DFOSC+gm5    &       & LTT 7987          \\
 20021001    &  52548.75      &  \phn+24.6          &  3500--7500     & A1.82+AFOSC+gm4    &       & BD+28 4211        \\
 20021010    &  52557.84      &  \phn+33.6          &  3600--9300     & INT+IDS+R150V      &       & Kopff 27          \\
 20030410    &  52739.17      &  +215.0             &  3500--8000     & TNG+LRB            &       & Feige 66          \\
\hline
\end{tabular}
\tablecomments{$^*$ Epoch relative to B$_{\mathrm{max}}$, which occurred on MJD=52524.2
\citep{gp:04}. A1.82 = Asiago 1.82\,m telescope, CA = Calar Alto 2.2\,m telescope, 
INT = Isaac Newton Telescope, D1.54 = Danish 1.54\,m telescope, NTT = ESO New Technology 
Telescope, TNG = Telescopio Nazionale Galileo. For the $-$8.3\,d spectrum we used the flux
standard taken with the same set-up on the previous night, while for the +4.6\,d spectrum
we used a sensitivity function appropriate for the setup, kindly provided by N. Elias de la Rosa.
$\dagger$ data not flatfielded; $\ddag$ missed target with blue grism.
             }
\end{centering}
   \end{table*}

\begin{table}
\caption[]{$E(B-V)$ values for the SNe shown in Fig. \ref{fig:compare}}
 \begin{centering}
 \begin{tabular}{llll}
 \label{tab:ebv}
                             &                 &                        \\
\hline
                             & $E(B-V)$       &  Reference     \\
\hline
 
 SN 1994D        & 0.06$\pm$0.02  &  \citet{patat:96} \\
 SN 1996X        & 0.1$\pm$0.03   &  \citet{salvo:01} \\
 SN 1998\,aq       & \nodata        &  \citet{branch:03} \\
 SN 1998bu       & 0.32$\pm$0.04  &  \citet{hernandez:00} \\
 SN 1999\,ee       & 0.32           &  \citet{hamuy:02} \\
 SN 2002bo       & 0.43$\pm$0.1   &  \citet{benetti:04a} \\
\hline
\end{tabular}
\tablecomments{The absence of narrow Na D absorption lines in SN 1998\,aq
               \citep{ay:98} implies little reddening.}
\end{centering}
   \end{table}

\section{Optical spectra at early times}

This section is subdivided as follows: we detail the early spectral
evolution of SN 2002er in Sect. \ref{sec:specevol} followed by a
discussion on the presence of carbon in Sect. \ref{sec:carbon}. 
We then describe the evolution of the expansion velocities in Sect. 
\ref{sec:vel}.

\subsection{Spectral Evolution}
\label{sec:specevol}

The spectral evolution of SN 2002er from day $-11$ to +34 is shown in
Fig. \ref{fig:specevol}. The rapid evolution at early times can be
followed in detail given the excellent temporal coverage.  The
outermost material in the ejecta, composed mainly of intermediate mass
elements is responsible for producing the spectrum up to about maximum
light.  The spectra shown in Fig. \ref{fig:specevol} are typical of
normal type~Ia supernovae, characterised at early epochs (up to about
day +6) by the deep P Cygni absorption near 6150\,{\AA} due to SiII
and the distinctive W-shaped SII feature near 5400\,{\AA}. Other
prominent features at these phases are attributed to the CaII H and K
blend, MgII $\lambda$\,4481\,{\AA} and a multitude of blends due to
FeII, SiII, and SII in the 4500 to 5000\,{\AA} range. The NaI~D lines
are clearly visible suggesting that substantial reddening due to
intervening material.

\begin{figure*}
 \begin{center}
 \mbox{
    \resizebox{0.52\textwidth}{!}{\includegraphics{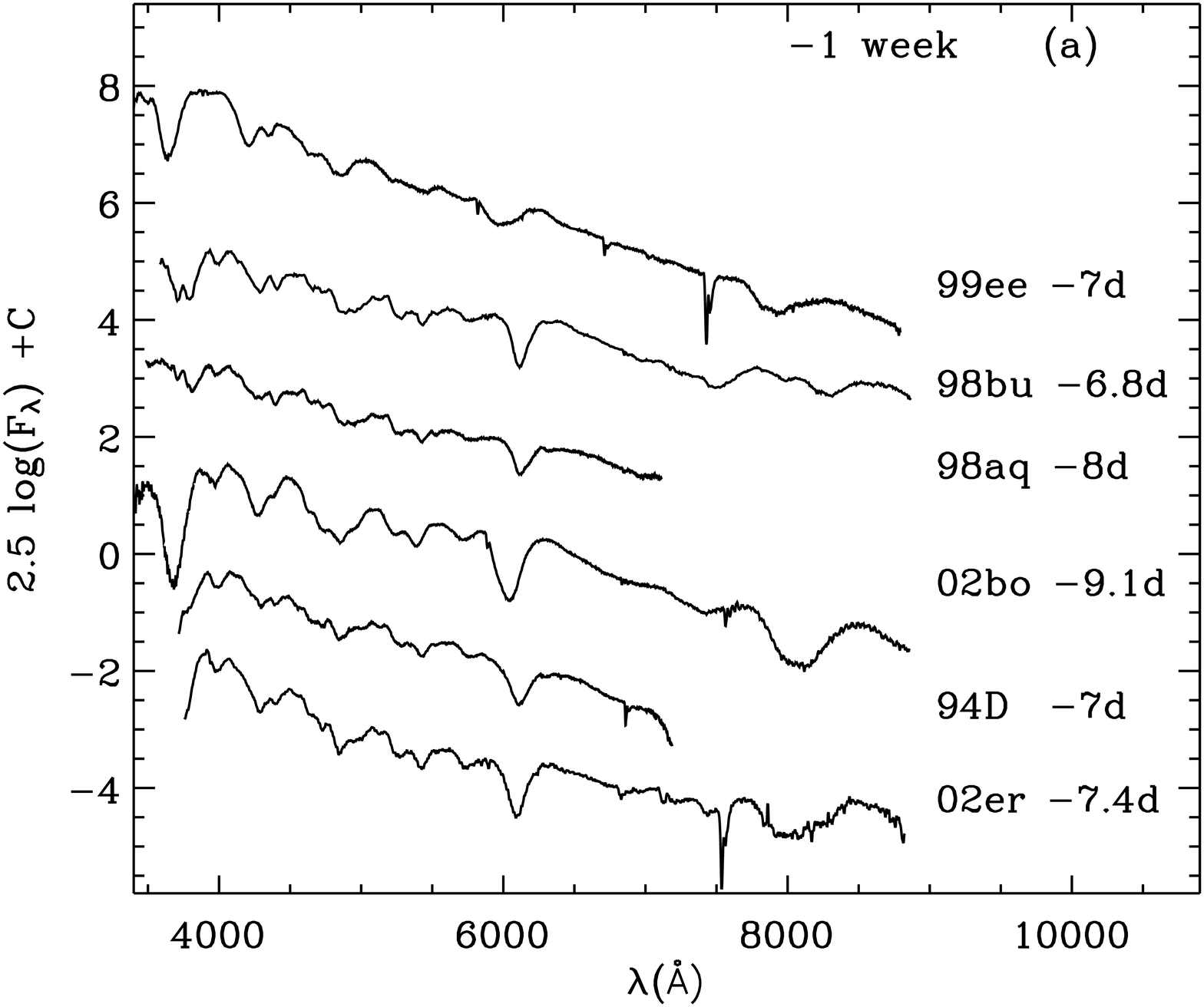}}
    \resizebox{0.52\textwidth}{!}{\includegraphics{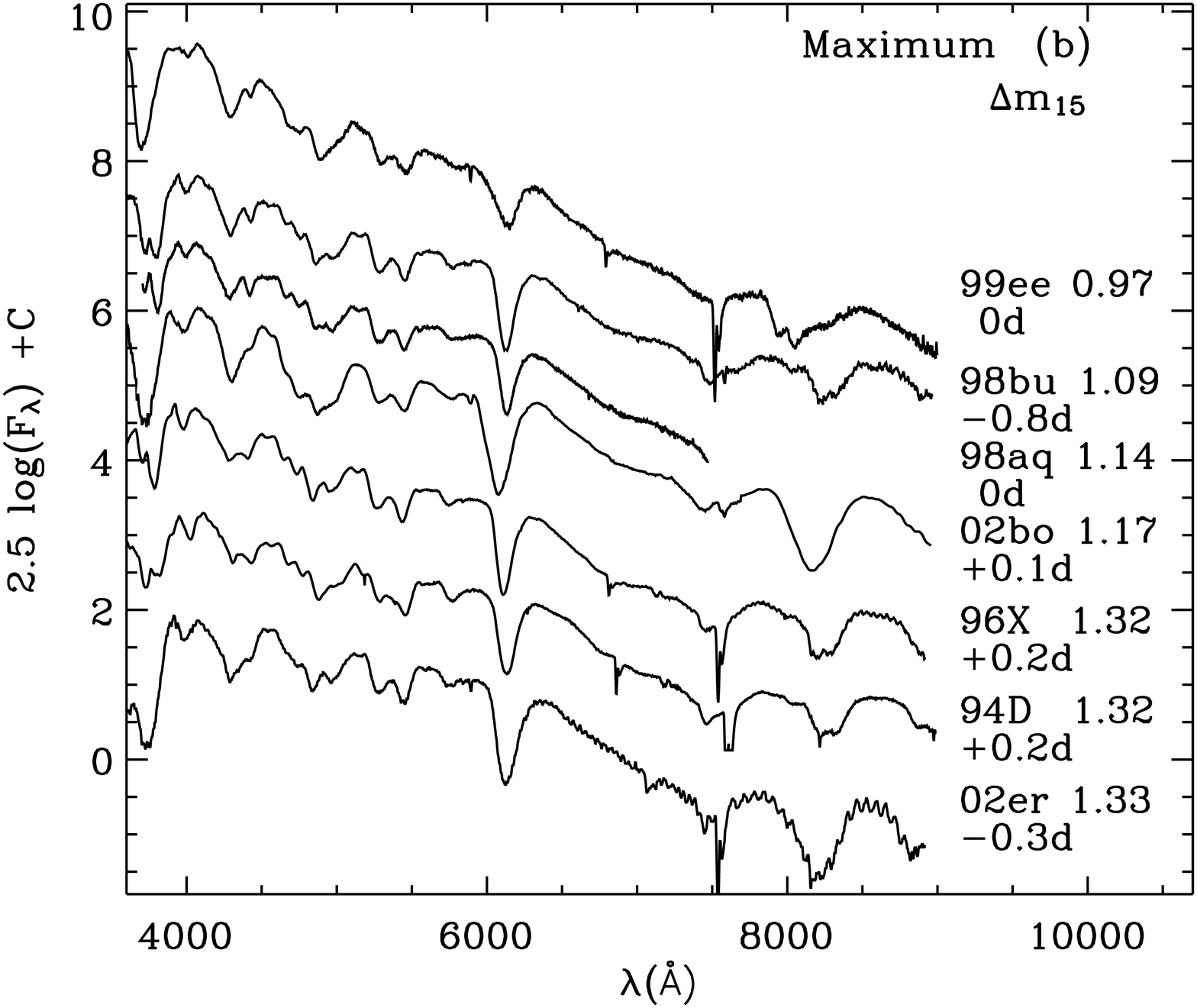}}
     }
\mbox{
    \resizebox{0.52\textwidth}{!}{\includegraphics{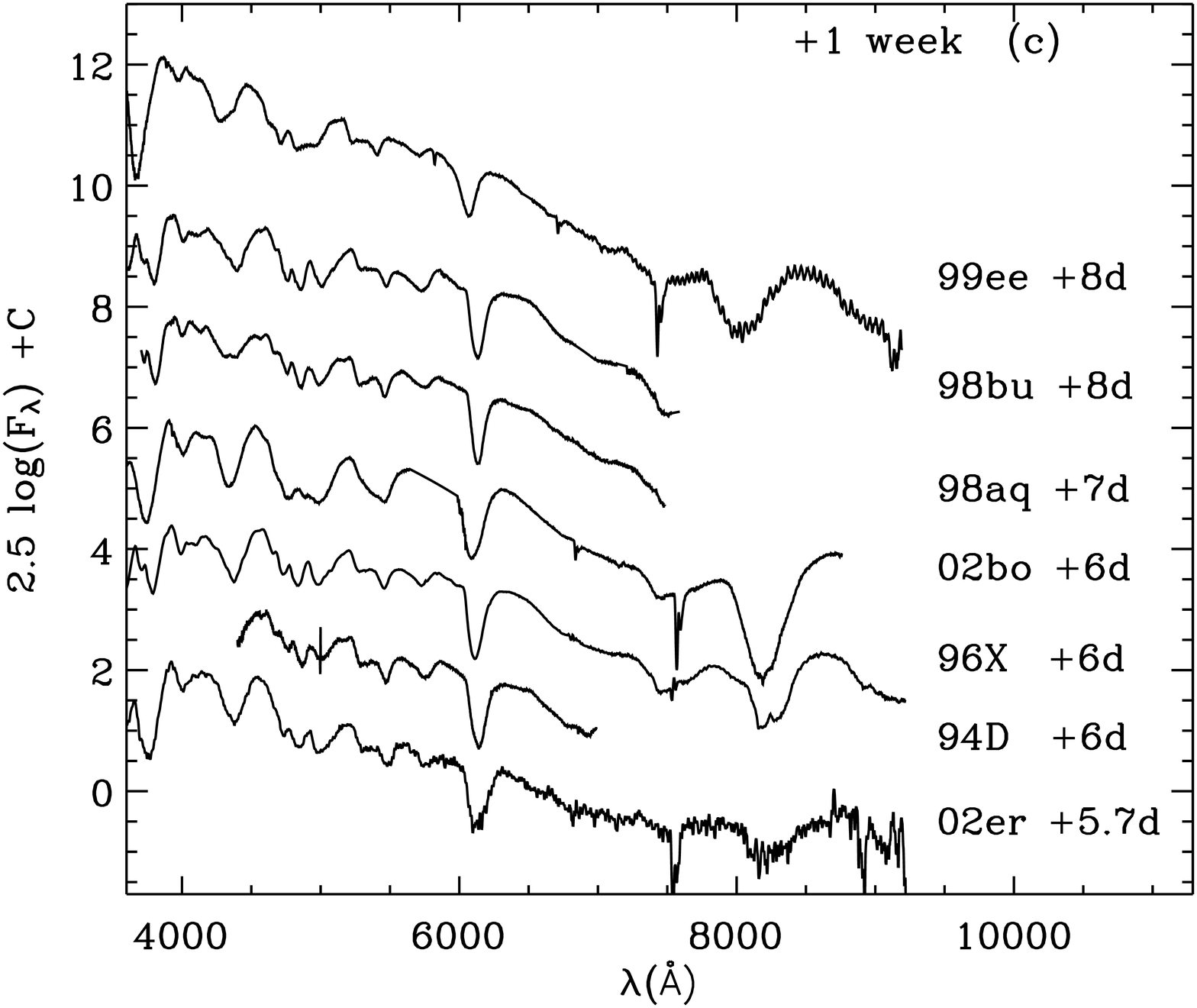}}
    \resizebox{0.52\textwidth}{!}{\includegraphics{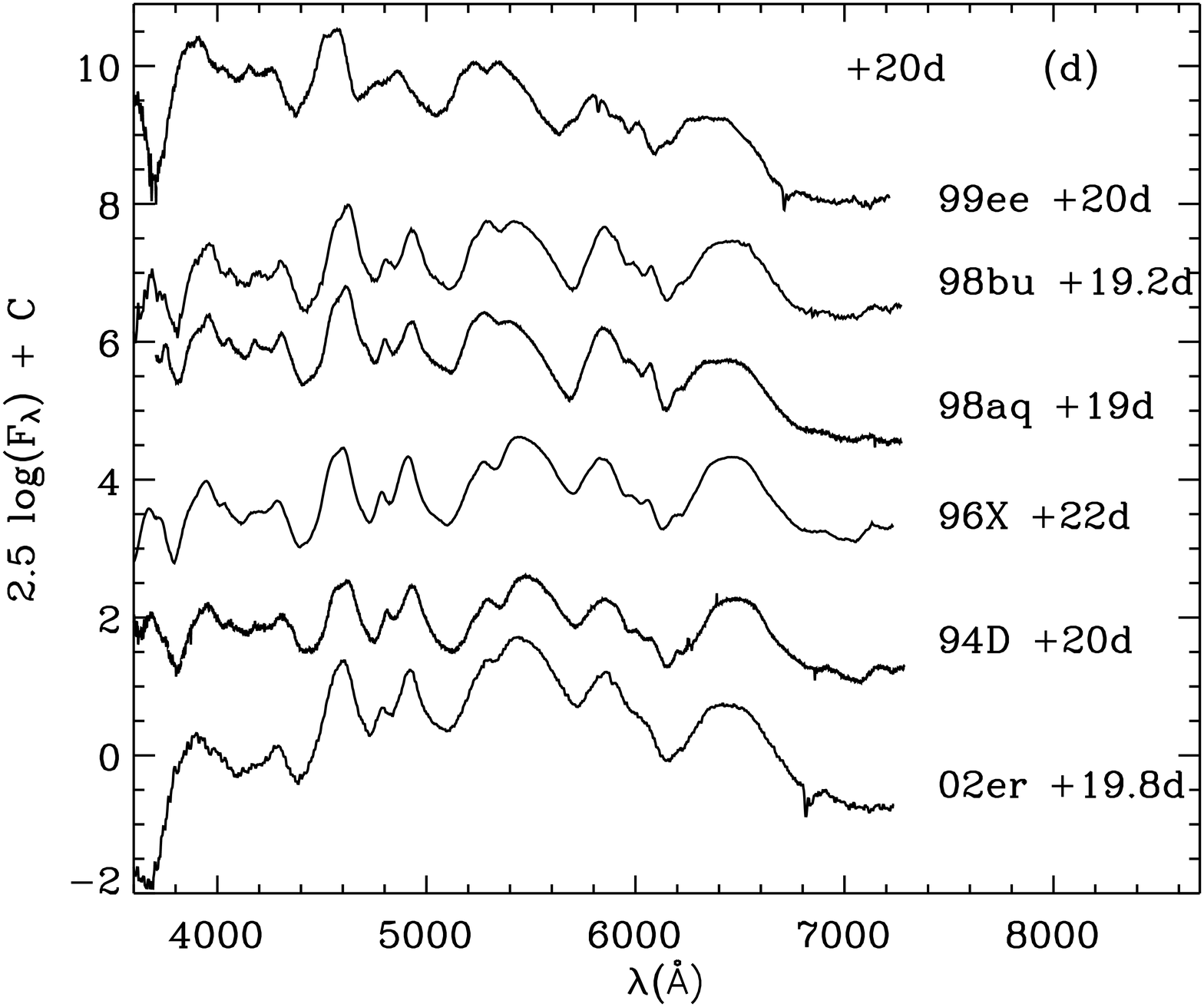}}
      }
\mbox{
    \resizebox{0.52\textwidth}{!}{\includegraphics{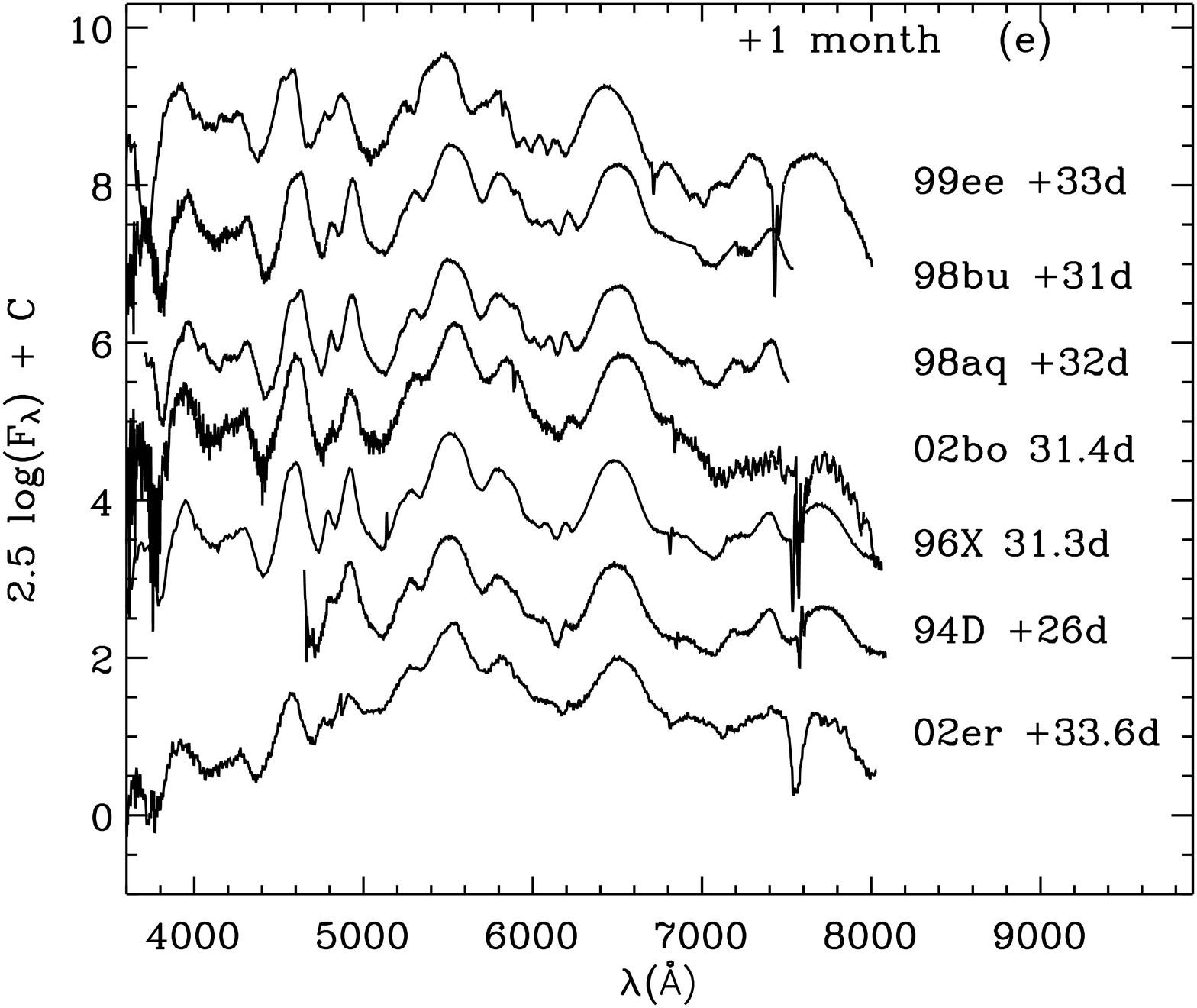}}
     } 
  \end{center}
\caption{Spectra of SN 2002er compared with other type Ia~SNe at four
representative epochs. The spectra have been shifted vertically by
arbitrary amounts and have been corrected for the respective redshifts
of the host galaxies. All spectra (except those of SN 1998aq) have
been corrected for extinction using the values in Table \ref{tab:ebv}
and the relation of \citet{cardelli:89}. The $\Delta m_{15}$ values
shown in panel (b) have been corrected for reddening following
\citet{phillips:99}.  Some of the principal spectral features at early
epochs are marked in panel (a).}
\label{fig:compare}
\end{figure*}

A comparison with other type~Ia supernovae at four representative
epochs is shown in Fig. \ref{fig:compare}. The decline-rate parameters
for these supernovae are fairly typical with the exception of the
slow-decliner SN 1999\,ee which shows markedly different behaviour at
early epochs, but has been included for comparison.  Although
differences between the spectra are present at all epochs, these are
most pronounced in the pre-maximum spectra (Fig. \ref{fig:compare}a).
The obvious contrasts are in the shape and depth of the prominent
features and in the amount of structure in the spectrum. The latter is
probably due in part to the differences in the velocities of the
ejecta; this is especially true of SN~2002bo for which the higher
velocities have probably smeared out the weaker features. However, the
similarity between SN~2002er and SN~1994D, down to the weakest of
features is striking. The strength of the temperature- sensitive SiIII
$\lambda$4553,4568\,{\AA} line suggests that SN~2002er and SN~1994D
spectra are intermediate in temperature compared with the hotter
SN~1998bu and SN~1998aq spectra and the cooler SN~2002bo spectrum.
The W-shaped SII feature at $\lambda$5454,5460\,{\AA} is already
apparent at $-11$\,d in the SN~2002er spectrum whereas it only became
apparent at $-$2\,d in the spectrum of SN~1999ee \citep{hamuy:02}.

By maximum light (Fig. \ref{fig:compare}b), the CaII lines near
8500\,{\AA} have become significantly stronger and the W-shaped SII
$\lambda$5454,5640\,{\AA} feature is very well-developed. The CaII H \&
K feature (blended with the SiII $\lambda$3858\,{\AA}) is more similar
to that of SN~2002bo than to the double-dipped feature seen in
SNe~1994D, 1996X and 1998bu. \citet{lentz:00} have suggested that the
presence or absence of a split in this feature is related to the
relative strength of the SiII and Ca features. The strength and width
of the SiII $\lambda$6355\,{\AA} feature is comparable to that seen in
SN~1994D and SN~1996X.  The $\lambda$4950\,{\AA} feature is stronger
in SN~2002er at pre-maximum epochs than in the other supernovae shown
in Fig. \ref{fig:compare}a; by maximum light, it has developed into a
pronounced W-shaped feature. Unlike the W-shaped SII feature at
$\sim$5400\,{\AA} which is ubiquitous at maximum in all normal type~Ia
supernovae, only in SN~1996X does the $\lambda$4950\,{\AA} feature
appear with comparable prominence to that of SN~2002er.  The higher
temperature of SN~2002er, as suggested by the SiIII lines, lends
credence to the suggestion that the $\lambda$4950\,{\AA} feature is
due to FeIII \citep{mazzali:95}. Note that the presence of lines of high
excitation is to be expected given the somewhat higher than normal
$^{56}$Ni mass inferred from the light curve by
\citep[0.7M$_\odot$][]{gp:04}.

\citet{nugent:95} found a tight correlation between the ratio, $\cal
R$(SiII) of the depths of the SiII $\lambda$5972\,{\AA} and
$\lambda$6355\,{\AA} absorption troughs near maximum light and the
luminosity of the supernova: the subluminous, faster decliners have
larger values of $\cal R$(SiII). However, this correlation has been
shown to break down for $\Delta m_{15}(B) \la 1.2$ \citep[see Fig. 8
in] []{benetti:04a}. For SN~2002er we measure 0.23 for $\cal R$(SiII)
which is consistent with the Nugent/Benetti relation to within the errors.

By about +20\,d post maximum light, the SiII $\lambda$6355\,{\AA}
feature is still clearly visible (Fig. \ref{fig:compare}d). All the
other supernovae, shown at comparable epochs, show kinks in the red
wing of the SiII $\lambda$6355\,{\AA} feature due to FeII.  SN~2002er,
on the other hand, shows a rather smooth red wing. This suggests that
the SiII feature is not confined to a narrow range in velocity and
corroborates the finding by \citet{gp:04} from light curve modelling,
that the ejecta are highly mixed. Notice also that the strength of the
CaII H\&K feature is more similar to that of the slowest decliner
shown in Fig. \ref{fig:compare}, SN 1999ee.  By about a month
post-maximum light (Fig. \ref{fig:compare}e), the SiII feature is
barely perceptible as features due to Fe dominate this wavelength
range. Figs.  \ref{fig:compare}d,e together suggest that the
intermediate-element phase was of slightly longer duration in
SN~2002er (i.e. slower transition to Fe-peak elements) than for the
other supernovae shown in Fig. \ref{fig:compare}.

Interestingly, although the similarity of SN~2002er with SN~1994D and
SN~1996X remains impressive throughout its photospheric evolution, the
differences become more apparent with time.

\subsection{Carbon in pre-maximum spectra}
\label{sec:carbon}

\begin{figure}[!t]
\plotone{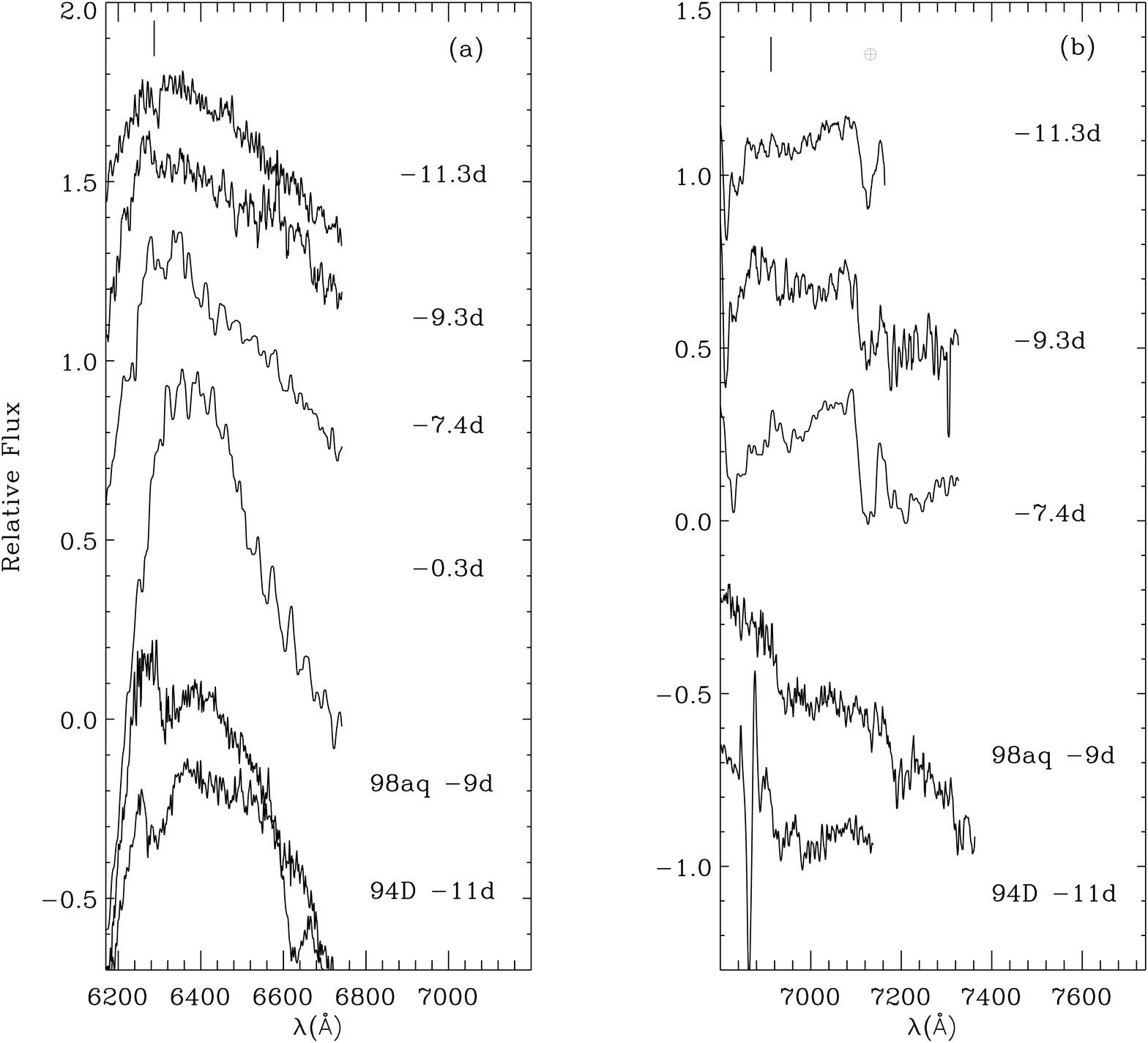}
\caption{Expanded view of several epochs of pre-maximum spectra around
the CII $\lambda$6580,7234\,{\AA} lines. The 02er spectra have been
smoothed and shifted vertically for clarity. \textbf{(a)} A short
vertical dash shows the position of the weak feature seen in the
earliest spectra. \textbf{(b)} The dash shows the location of the
$\lambda$7234\,{\AA} line at the velocity (13000\,\kms) inferred from
panel (a). The earliest spectra of SN~1994D and SN~1998aq are shown
for comparison.}
\label{fig:carbon}
\end{figure}

\begin{figure}[!t]
\plotone{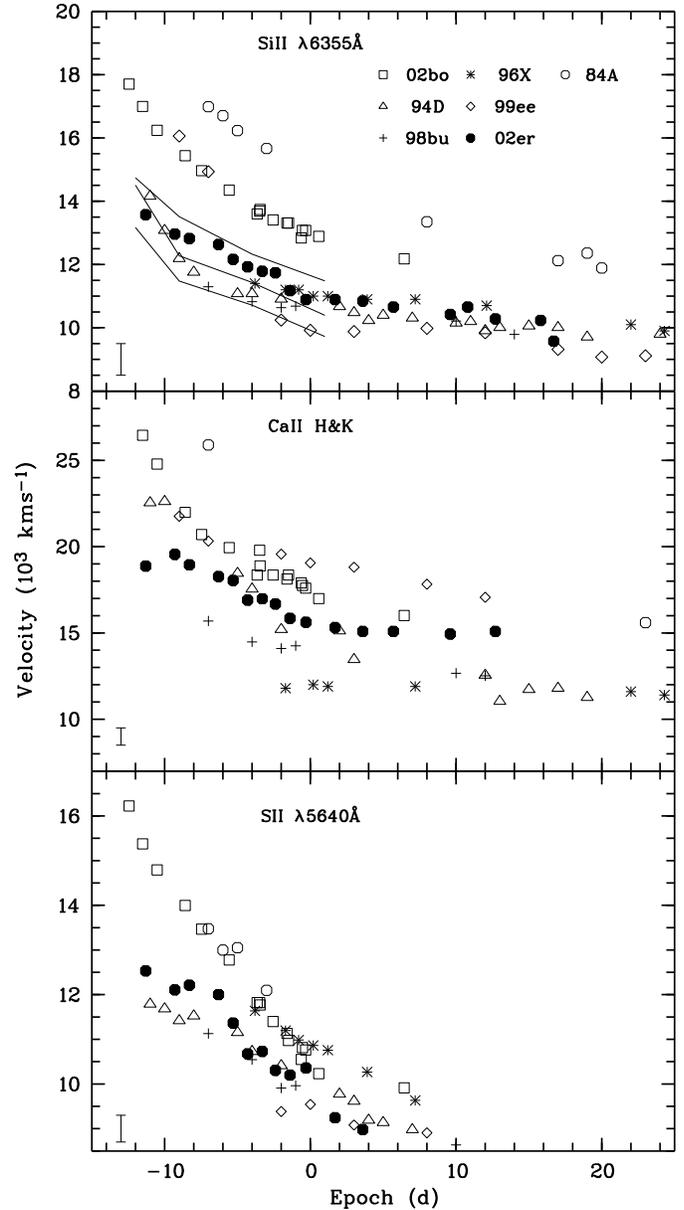}  
\caption{The velocity evolution of the SiII $\lambda$6355\,{\AA}, Ca
H\,\& K, and SII $\lambda$5640\,{\AA} features compared with other SNe
as measured from the minima of the respective absorption features. The
solid curves shown in the top panel are models of different
metallicity (top: $\times$10 middle: $\times$1 bottom:$\times$1/10
solar) obtained from \citet{lentz:00}. Typical error bars for 
SN~2002er are shown in the bottom left-hand corner of each panel.}
\label{fig:vel}
\end{figure}

Carbon is a diagnostically interesting element in pre-maximum spectra.
Its abundance allows us to place limits on the composition of the
progenitor white dwarf.  In addition, the amount of carbon constrains
the extent of burning in the outer layers thereby testing various
explosion scenarios.  For instance, current delayed-detonation models
have very little carbon below about 30000\,\kms, while the 1-D
deflagration model (W7) has unburnt carbon present down to velocities
of about 14000\,\kms \citep[e.g.][]{branch:03}.

The first reported detection of carbon in a SN~Ia was in the $-$14\,d
spectrum of SN~1990N \citep{fisher:97}. However, carbon lines have
also been identified at $-$10\,d in the spectra of SN~1994D and at
$-$9\,d in SN~1998aq \citep{branch:03}.  The strongest line is at
$\lambda$6580\,{\AA} with weaker lines at $\lambda$4267\,{\AA} and
$\lambda$7234\,{\AA}.  In Fig. \ref{fig:carbon}, we show tentative
evidence for the presence of the CII $\lambda$6580\,{\AA} line present
up to about a week before maximum light at about 13000\,\kms.  There
is no evidence of the CII $\lambda$7234\,{\AA} line
(Fig. \ref{fig:carbon}b) but this would, in any case, be expected to
be much weaker.

\subsection{Expansion velocities}
\label{sec:vel}

As the photosphere recedes into deeper and slower moving material, the
minima of the absorption features shift redwards with time.  As
already mentioned, this can provide a powerful ``tomographic'' picture
of the SN ejecta.  However, very strong lines may be less useful in
this regard.  \citet{patat:96} point out that the velocities inferred
from the strong features (e.g. the characteristic SiII line) are only
representative of the photospheric velocity at early phases when the
amount of material above the photosphere is small; as the photosphere
retreats, it encounters flatter density gradients which results in
strong lines being formed over a large velocity range, thus
overestimating the photospheric velocity. In contrast, weak lines can
provide good indicators of the photospheric position.

In SN~2002er, we examined the velocity evolution of the strong SiII
$\lambda$6355\,{\AA} and CaII H\&K lines and the weak SII
$\lambda$5640\,{\AA} line. The results are shown in Fig. \ref{fig:vel}
together with those of other supernovae for comparison.  All
velocities have been corrected for the recession velocities of the
respective host galaxies.  As expected, the weak SII line exhibits
lower velocities than do stronger lines.  The strong SiII
feature in the earliest spectrum of SN 2002er ($-$11.3\,d), has its
minimum centred at about 6120\,{\AA}, with its blue edge lying at
about 22000\,\kms. As has already been pointed out by several authors
\citep[e.g.][]{mazzali:93} such velocities are well in excess of the
15000\,\kms\ limit in the unmixed W7 model \citep{nomoto:84}.

\citet{lentz:00} have computed the effects of varying the metallicity
in the C+O layer of the standard deflagration model, W7, on the
emergent pre-maximum spectra of type~Ia supernovae. They find that the
strength, profile, and velocity of the SiII $\lambda$6355\,{\AA}
feature is a function of metallicity. Their results are plotted in
Fig. \ref{fig:vel}.  Clearly the range of velocities seen between a
number of SNe~Ia may indeed be due to
metallicity variations.  However, even with implausibly large
variations in the C+O layer metallicity ($\times$1/10 to $\times$10
solar) we cannot reproduce the high velocities exhibited by SN~2002bo
and SN~1984A.  These large velocities could be due to other factors
such as higher explosion energy.

\section{Synthetic spectra}

Comparison with observations of synthetic spectra based on specific
explosion models provides a powerful tool for probing the physics of
supernovae.  Here we compare the SN~2002er early-time spectra with
model spectra generated by a Monte Carlo (MC) code. The code is based
on the Sobolev approximation, and it was discussed in a number of
papers that the reader can turn to for details
\citep{al85,ml93,l99,m00}.  The input parameters are the luminosity
$L$, the epoch $t_{exp}$ where explosion occurs at $t_{exp}=0$\,d , the
photospheric velocity $v_{ph}$, and a density distribution from an
explosion model. The code iteratively computes the radiation field and
the level populations in the SN ejecta under the assumption that there
is no net exchange of energy between matter and radiation above a
sharply defined photosphere. This means that no deposition of fast
particles created in radioactive decays occurs above the photosphere,
and that absorptive continuum opacities are ignored. This is a
reasonable approximation at the earliest phases of a SN~Ia.

The luminosity is chosen to match the observed value and the velocity
adjusted to obtain a good match to the observed line blueshifts. The
epoch relative to explosion is determined, inevitably with some
uncertainty, from the time of observed maximum. This is because first
observations are usually at least a few days after the SN exploded.
Different models can be chosen for the density distribution. In this
paper we base our calculations on W7\citep{nomoto:85}, which is known
to give a good description of typical type Ia SNe.  Energy packets are
released at the photosphere and travel through the SN envelope, where
they can undergo line absorption and reemission (branching) and
electron scattering. When a converged temperature structure is
established and appropriate excitation and ionisation conditions are
computed (using a modified nebular approximation \citep{al85,ml93}), a
synthetic spectrum is obtained using a formal integral scheme.

We calculated two spectral models for SN~2002er, one for an epoch near
maximum light (day --2.4) and one for the earliest epoch at --11.3
days. These two epochs were chosen to cover a period in which the
physical properties change significantly.

In the first iteration of the day --2.4 spectral model we adopted a
typical value for the photospheric velocity at this epoch, of
9500\kms. The distance modulus ($\mu=32.9$) and the reddening
($E(B-V)= 0.36$) were taken from \citet{gp:04}.  This initial set of
parameters provided a poor match to the observed spectrum.  The
problem was that, in order to match the observed bolometric luminosity
of log$_{10}L \approx $ 43.2 [erg\,s$^{-1}$] \citep{gp:04} and the
observed line velocity (this work) the model temperature near the
photosphere had the rather high value of $T_{BB}\approx 14,000$\,K.
Consequently the ionisation was much too high.  For example, strong
\SiIII\ lines were predicted but not observed.  The large observed
line ratio of \SiII\ 5970\Ang\ to \SiII\ 6350\Ang\ also
indicates a rather low temperature. \citet{nugent:95} derived a
temperature of $\sim 9000$\,K for SNe~Ia with a line ratio similar to
that of SN~2002er.

\begin{figure}[!t]
\includegraphics[width=.33\textwidth,angle=-90]{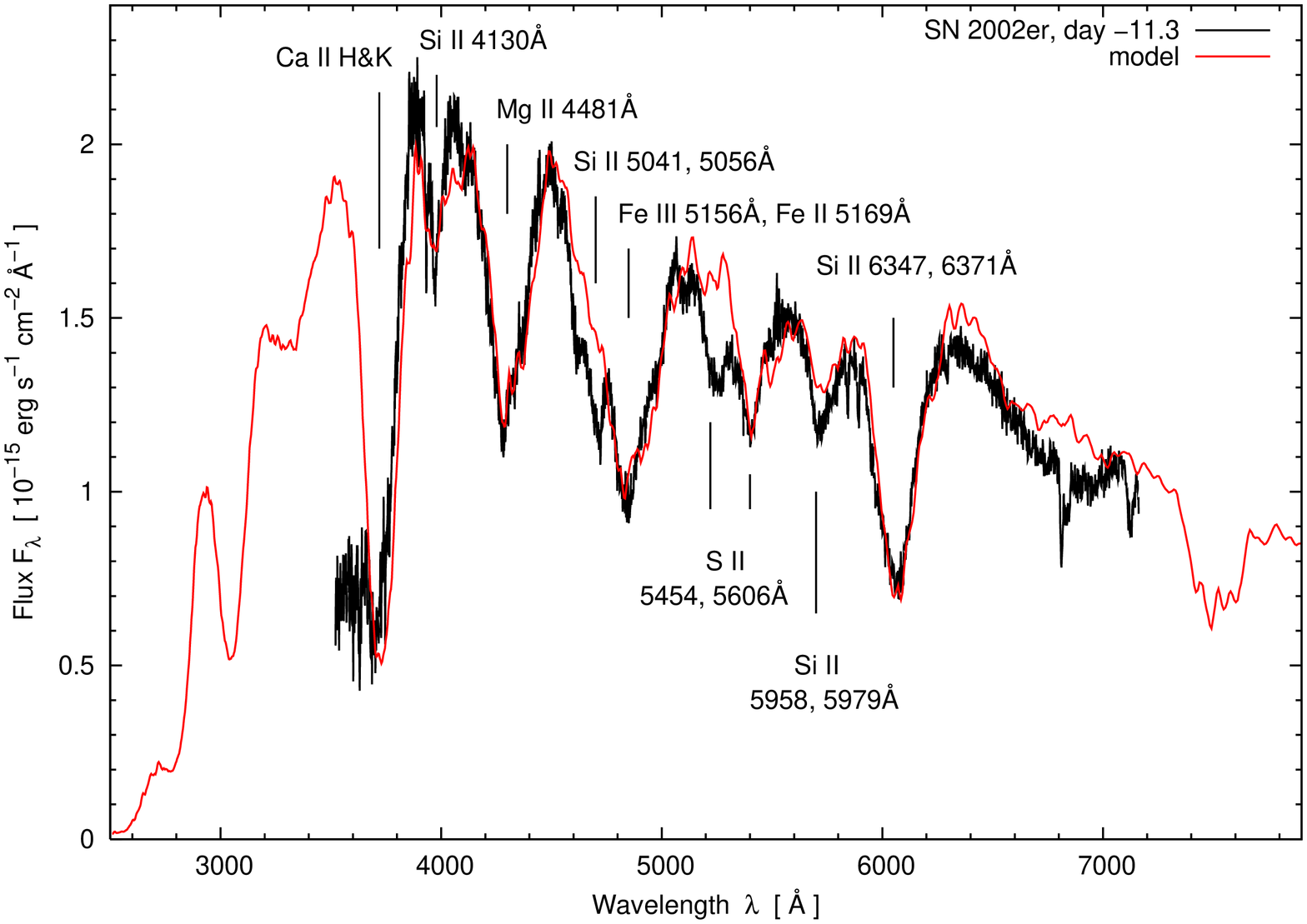}
\includegraphics[width=.33\textwidth,angle=-90]{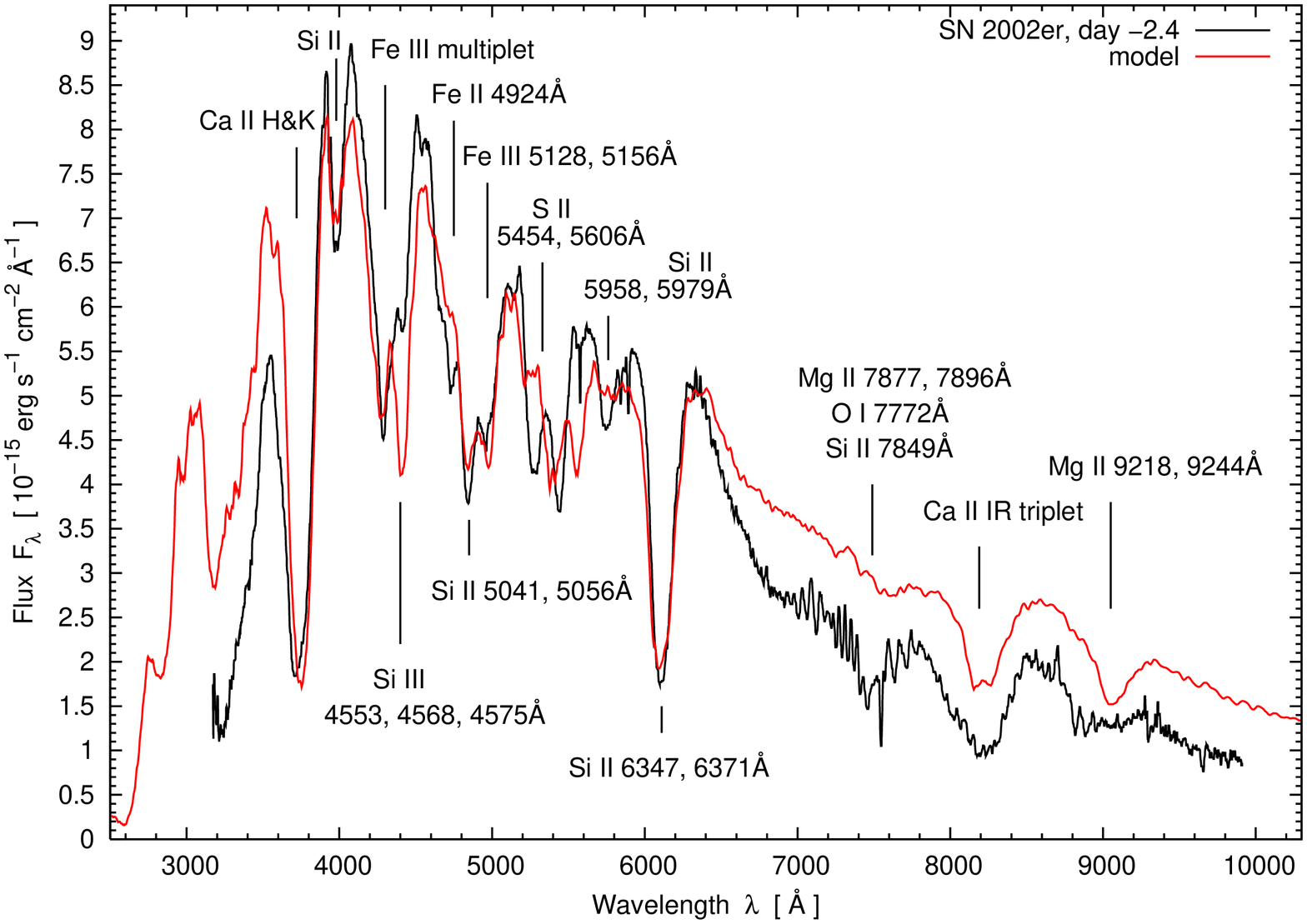}
\caption{Observed spectrum and corresponding model of SN~2002er at 
using $E(B-V)=0.36$ and $\mu = 32.70$ with  $t_{\mathrm{exp}} = 8.7$\,d
(top panel) and  $t_{\mathrm{exp}}=17.6$\,d (bottom panel). 
}
\label{-11.3}
\end{figure}

In order to try to resolve this discrepancy between the model and
observations, we explored the range of parameter space available
through the uncertainties in reddening, distance and explosion
epoch. We found that even if we set the reddening at the 1~$\sigma$
lower limit of $E(B-V)=0.31$ \citep{gp:04}, we make little difference
to the discrepancy. The best match was obtained by pushing the
distance modulus to its 1$\sigma$ lower limit of $\mu = 32.7$, while
keeping the formal value of the reddening, $E(B-V)=0.36$, and using a
risetime of 20.0d.  This risetime is 1.3d longer than estimated by
\citet{gp:04}, but is closer to the typical SN~Ia risetime of
$t_{B_{\mathrm{max}}} = 19.5$\,d \citep{ri99}. This departure from the
\citet{gp:04} value can be justified by the fact that the photometric
data are not available earlier than --7.3 days.  Nevertheless, even
with these more extreme parameter values, the temperature was still
found to be rather too high.  The best match parameters are log$_{10}L
= 43.15$[\ergs] and a velocity at photophere of 9200\kms, leading to a
photospheric temperature of $T_{\mathrm{BB}} = 12,460$\,K.  The model
thus obtained is compared with the observed spectrum in
Fig.~\ref{-11.3}b.

The main features are labelled with their rest wavelength in
Fig.~\ref{-11.3}.  We identify strong \SiII\ lines at 6360\Ang,
5970\Ang, and at 4130\Ang. Additionally, the deep absorption near
4800\Ang\ contains a significant amount of \SiII. The rest of this
feature is caused by \FeII. Weak \SiIII\ lines can be seen in the
observed spectrum at 4570\Ang\ and 5740\Ang, but they are both very
strong in the model due to the high temperature. \SII\ produces the
double line feature near 5400\Ang. The absorption at $\sim$\,4400\Ang\
is due to \MgII\ 4481\Ang, as well as the one near 7500\Ang.  Finally
we recognise the strong \CaII~H\&K lines and the \CaII\ IR triplet.
Although it is not as pronounced as in the case of SN~2002bo
\citep{stehle:04}, a high-velocity component in the \CaII\ lines can
be seen that is not reproduced by the model, which uses the density
distribution of W7.

In constructing the spectral model for --11.3 days, we could not
obtain the flux directly from observation since the earliest
photometry epoch was at --7.3 days.  [One possibility is to
extrapolate the light curve to the early epochs using a quadratic
function in time \citep{riess:99} ($L \propto t_{exp}^2$). This
procedure gives log$_{10}L \simeq 42.75$[\ergs]. This is however too
bright, leading to a very hot model.  
An alternative approach is to used the synthetic light curve fit of
\citet{gp:04} to extrapolate back to --11.3 days. This yielded
log$_{10}L \simeq 42.6$[\ergs].  This allowed the model to produce much
more reasonable results. The best match was obtained with a velocity
at the photosphere of 13000\kms, leading to $T_{BB}=10950$\,K. The
model spectrum is compared with the observed spectrum in
Fig.~\ref{-11.3}.  Although the agreement is good, it should be kept
in mind that the synthetic light curve shape may change when a more
realistic element distribution is adopted compared with that used in
\citet{gp:04}.  It is interesting that in SN~2002er, as in SN~2002bo,
the temperature {\it rises} as we approach maximum light. This
behaviour is not typical of all SNe~Ia \citep{benetti:04b}

While the strong \SiII\ line absorptions are well reproduced, \SiII\
5970\Ang\ is too weak owing to the temperature still being too
high. The \SII\ double line feature near 5400\Ang\ is not well
reproduced in its blue part. Since these lines come from the same
lower levels, there may be a problem in the model line list, in
particular the $gf$ values. The line at $\sim$\,4700\Ang\ is not
reproduced by the model. This feature was detected in other SNe~Ia and
can possibly be assigned to a high-velocity absorption of \FeII\
\citep{ha99}. All the other lines are well reproduced and have already
been discussed for the day --2.4 model.

The above analysis indicates shortcomings in the modelling procedure.
Specifically, owing to the homogenous abundance distribution of the
model, to reproduce the strong \SiII\ absorptions we had to introduce
a rather high Si abundance (63\% by mass) throughout the envelope. But
due to the high photospheric temperature, the Si is almost completely
doubly-ionised near the photosphere at maximum, severely overproducing
the \SiIII\ features.  Only further out in the envelope, where the
temperature is sufficiently low, are the required \SiII\ absorptions
formed.  A similar problem was encountered with SN~2002bo
\citep{benetti:04a}.  However, \citet{stehle:04} find that, in a more
recent analysis of SN~2002bo by the ``Abundance Tomography'' method,
it is possible to reduce the Si abundance near the photosphere and yet
simultaneously keep a desirably high Si abundance at the outer radii
\citet{stehle:04}.  Furthermore, \citet{stehle:04} also find that
improving the description of the radiation field at the inner boundary
reduces the overestimate of the emergent flux in the red, leading to a
decrease of the total luminosity and hence a lower temperature near
the photosphere.  In the case of SN~2002er, however, we also note that
if the reddening and/or the distance are reduced by more than the
1\,$\sigma$ error boundaries estimated by \citep{gp:04} then greater
consistency with the spectral models is achieved.  At present, there
is no independent evidence that such reduction should be applied.

We conclude that the maximum light spectrum of SN~2002er can be fairly
well modelled using a homogeneous abundance distribution in the
atmosphere provided we push the adopted distance and risetime close to
the observational limits.  Modelling of the earliest spectrum depends
sensitively on the adopted flux level, but unfortunately observations
do not directly provide this value.  However, extrapolation by means
of a bolometric light curve model fit appears to provide a reasonable
estimate of this quantity.  Future progress will require a more
realistic model which includes a stratified abundance distribution.

\begin{figure}
\plotone{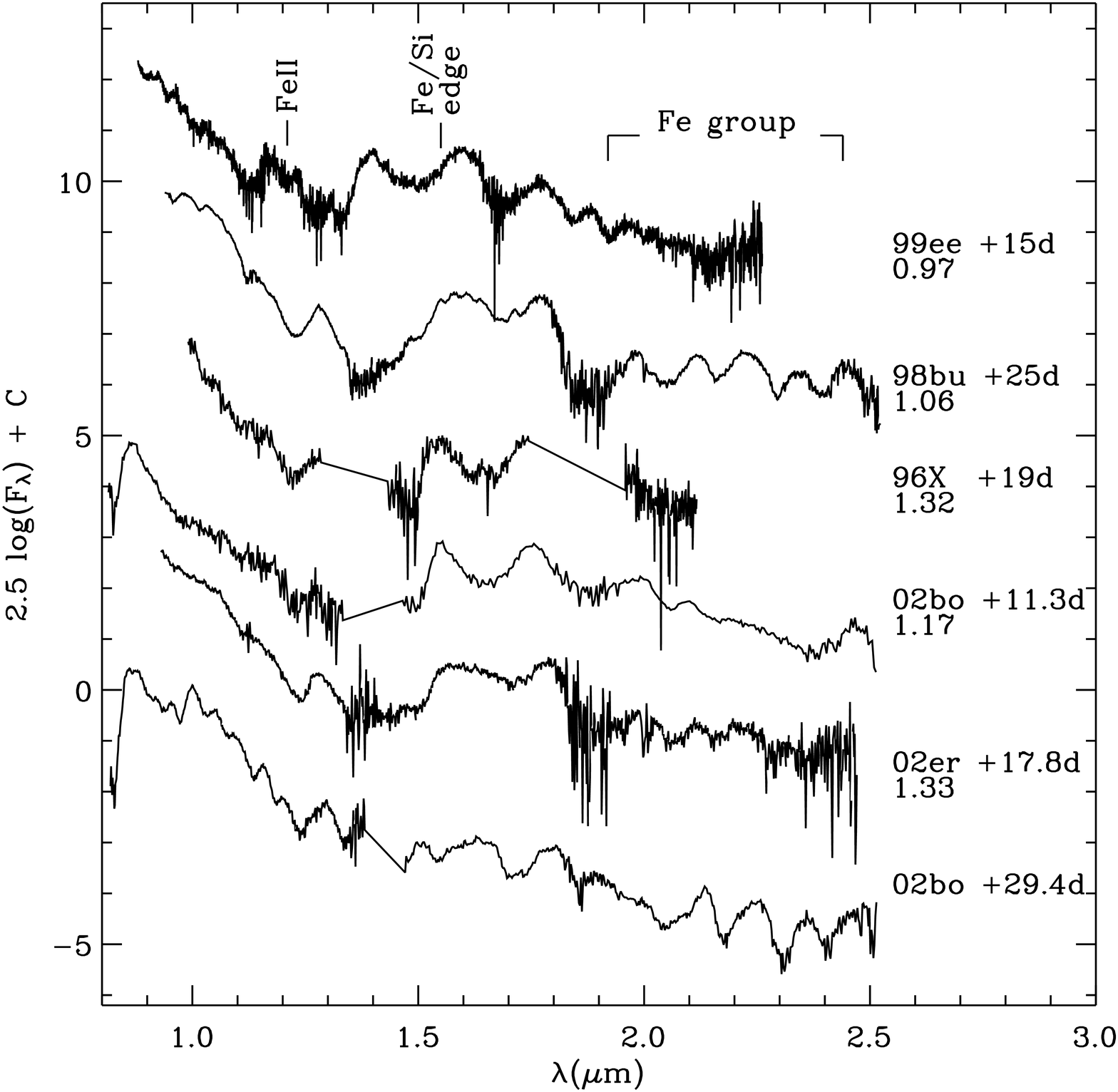}
\caption{Near-infrared spectrum of SN~2002er at +18\,d compared with
spectra of other SNe~Ia at comparable epochs. The spectra of
SNe~1998bu, 1996X, 1999ee, and 2002bo have been taken from \citet{jha:99},
\citet{hernandez:00}, \citet{hamuy:02}, and \citet{benetti:04a} respectively. 
The spectra are shown in the respective rest-frames of the supernovae and have 
been de-reddened using the values in Table \ref{tab:ebv}. 
The epochs relative to maximum light are shown
next to the name and the $\Delta m_{15}$ values are noted just
below. Prominent features are indicated following \citet{marion:03}.}
\label{fig:nirspec}
\end{figure}

\begin{figure}
\plotone{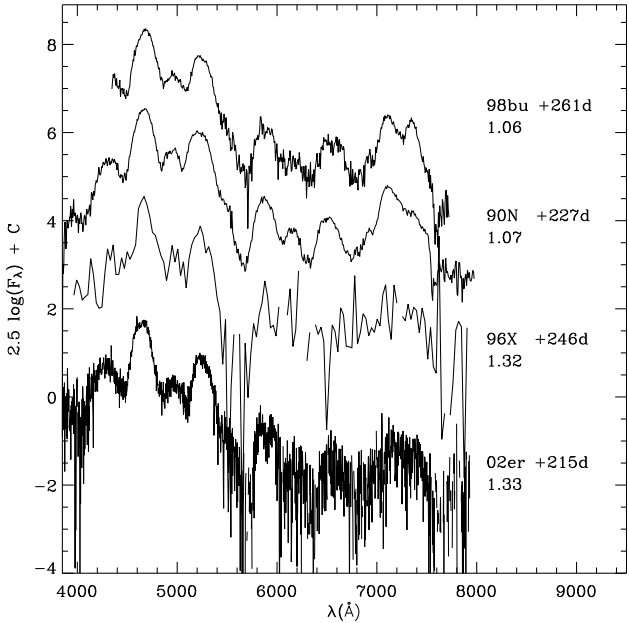}
\caption{The nebular spectrum of SN~2002er compared with the few 
spectra that are available of other SNe~Ia at comparable
epochs.  The epochs relative to maximum light are indicated next to
the names and the $\Delta m_{15}$ values are indicated just below. The
spectra have been corrected for the redshift of their respective host
galaxies and have been dereddened in the same manner as for
Fig. \ref{fig:compare}.  $F_{\lambda}$ is in units of 10$^{-16}\,$erg
cm$^{-1}$s$^{-1}${\AA}$^{-1}$.  The SN 1998bu, SN 1990N, and SN 1996X spectra
have been shifted by +5.2, +4.0, and +3.2 respectively. The SN 1996X spectrum
has been binned by a factor of 10 in an attempt to compensate for the very
low signal in the 6000-7000\,{\AA} region.}
\label{fig:neb}
\end{figure}

\section{Near-IR spectroscopy of SN 2002er}

Early-time IR spectroscopic observations can provide a particularly
powerful diagnostic of the dynamic model because they probe different
depths at the same epoch within the exploded white dwarf through the
strongly variable line-blanketing opacity \citep{wheeler:98}.
Furthermore, the large number of lines in the optical/UV regions,
coupled with the strong Doppler-broadening (several thousand
kms$^{-1}$) can make species identification difficult. Strong
extinction effects often introduce additional error.  In contrast, the
near-IR has fewer lines and reduced sensitivity to extinction
uncertainty, allowing firm line identification and accurate
measurement of line strength and evolution.  In addition, since there
are much fewer strong resonance lines, it means that, in general, for
a line to be detectable it needs a significant element abundance. In
other words, IR lines suffer less from contamination by primordial
elements. 

After maximum light there is a rich mixture of Fe-group lines in the
IR spectrum, allowing us to explore the innermost regions of the
supernova and to ascertain the extent of incomplete silicon burning.
\citet{wheeler:98} find that the post-maximum-light features at
$\sim$1.5--1.7 and 2.2--2.6\,$\mu$m are primarily due to iron group
elements of the second ionisation stage.  The blue extent of the
feature at $\sim$1.55\,$\mu$m defines the transition from partial to
complete Si burning \citep{marion:03}.  Due to the high optical depth,
this feature is formed near the outer boundary of the Fe/Co/Ni core
and demarcates the the Fe/Si edge (Fig. \ref{fig:nirspec})
\citet{wheeler:98} particularly note that the minimum at
$\sim$1.7\,$\mu$m provides a diagnostic tool for assessing the
velocity spread of the region containing radioactive materials which,
in turn, depends on the underlying explosion scenario.  The
2.2--2.6\,$\mu$m features also contain significant contributions from
intermediate mass elements such as Si. Indeed, these K-band features
are formed in the transition region between complete and incomplete
silicon burning and are sensitive to the composition structure and
hence the explosion model.

Due to a combination of poor conditions and instrument availability,
only a single epoch of near-IR spectroscopy was obtained, on 2002
September 24 (+17.8\,d) using the NTT/SofI low-resolution blue and red
grisms. 
The spectrum was reduced using standard procedures in the
FIGARO 4 environment. Wavelength calibration of the 0.95-1.64\,$\mu$m
spectrum was carried out using a Xe arc; as no arc frame was available
for the 1.53-2.52\,$\mu$m spectrum, we used the OH sky lines. Flux
calibration was carried out with respect to the bright F6V star
HIP\,95550 and compared with near-IR photometry taken at +19.9\,d
\citep{gp:04}. The spectrum, scaled by 20\% to match the photometry,
is shown in Fig. \ref{fig:nirspec}.  For comparison, we also show the
rest-frame near-IR spectra of other SNe~Ia at comparable epochs.  
Owing to the sparsity of near-IR spectra at the appropriate epoch
($\sim$3~weeks), we include two spectra of SN 2002bo that bracket
the phase of the SN 2002er near-IR observation.

Inspection of Fig. \ref{fig:nirspec} shows that the IR spectrum of
SN~2002er is fairly typical for the epoch considered.  As anticipated
for this epoch, the spectrum is dominated by the increasingly exposed
Fe-group elements.  The 1.7$\mu$m feature is considerably shallower
in SN~2002er than in SN~1996X and is more reminiscent of the
morphology seen in the somewhat faster decliners such as SN~1998bu
and perhaps, SN~1999ee.  As indicated above, this may be interpreted
as a larger-than-average velocity spread in the radioactive
materials. However, the measured Doppler blueshift in the 1.5$\mu$m
edge \citep[rest wavelength 1.57$\mu$m,][]{marion:03} is
8300$\pm$500\,\kms\ for SN~2002er compared with 10500$\pm$500\,\kms\
for SN 1996X at +19d. This suggests that silicon burning actually extended 
to higher velocities in SN~1996X.

\section{The nebular spectrum at +215\,days}

In a Type~Ia supernova, by about 60--100 days post-explosion the
photosphere has receded to the centre, and the ejecta have become
transparent to the optical/IR emission i.e. the
forbidden-line-dominated nebular phase has been reached. The
transparency of the ejecta means that the physical conditions and
constituents of the innermost regions can be directly
probed. Furthermore homologous expansion means that line profiles
provide valuable information about the distribution of the
nucleosynthesised materials. In particular, the mass and velocity
distribution of $^{56}$Ni and its products can provide information on
the nature of the explosion.

We obtained a single nebular phase spectrum at +215\,d post
$B$-maximum.  The spectrum was reduced as described in
Sec. \ref{sec:obs} and is shown in Fig. \ref{fig:neb} together with
spectra of other type Ia supernova at comparable epochs.  As with the
earlier epoch spectra, the form of the late time spectrum of SN~2002er
is not unusual.  The spectrum is dominated by forbidden lines of
singly and doubly ionised Co and Fe. The FWHM of the 4700\,{\AA}
feature ($\sim$13000\,\kms) is consistent with the relation in
\citet{mazzali:98} that links the width of nebular lines to the
luminosity of the supernova.

\begin{figure}
\plotone{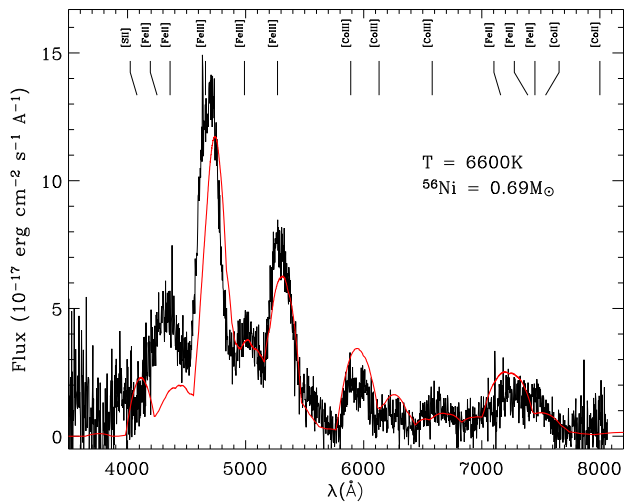}
\caption{The nebular spectrum of SN~2002er overlaid with a one-zone
         model. The model has been redshifted and reddened using the
         recession velocity of the host galaxy and $E(B-V)=0.36$. The
         main features have been labelled; note that most of these 
         features are blends of several lines.}
\label{fig:onezonemod}
\end{figure}

In order to exploit the information available in the nebular spectrum,
we compared it with the non-LTE nebular spectral model described in
\citet{bowers:97}, to which we refer the reader for details.  Briefly,
the model comprises a uniform density, homologous expanding sphere
containing iron, cobalt, and sulphur with the relative abundances as
specified by radioactive decay. Forbidden lines of these elements
emitted by singly and doubly-ionised species are predicted to dominate
the nebular phase at optical and near-IR wavelengths. Consequently
only these two ionisation stages were included in the model.
The distance and reddening were set at $\mu=32.9$ and $E(B-V)=0.36$
respectively (see above).  The $^{56}$Ni mass, expansion velocity,
ionisation degree and temperature are free parameters.  The best-match
model is shown in Fig. \ref{fig:onezonemod}, compared with the
observed spectrum. The model parameter values are $T=6600$\,K, 
expansion velocity~$\sim$9000\kms\, fractional abundance of SII=0.5,
and M($^{56}$Ni)= 0.69\,$M_\odot$.  These clearly yield a fair match
to the observed spectrum.  We also note that the derived mass of 
$^{56}$Ni is close to the value obtained from the bolometric light
curve \citep[0.7\,$M_\odot$,][]{gp:04}.  We conclude that (i) the
masses of $^{56}$Ni derived by the two methods are consistent with
each other, and (ii) the fraction of neutral and triply-ionised
iron-group elements is indeed quite small.

\section{Summary}

We have presented extensive optical spectroscopy of SN~2002er at
epochs spanning $-$11 to +215\,days. The spectroscopic coverage at
epochs up to +34\,days is especially noteworthy. In most respects,
SN~2002er bears the hallmarks of being a typical, albeit significantly
reddened, type Ia supernova.  Its spectral evolution was similar to
those of other normal type Ia SNe such as SN~1994D and SN~1996X that
suffered from minimal reddening. Nevertheless, differences are seen
between the coeval spectra of SN~2002er and other normal SNe~Ia. These
are most pronounced in the pre-maximum spectra.

We compared the spectra at two early-time epochs with a homogeneous
spectral synthesis model based on the deflagration model W7.  The best
match was obtained only by pushing the distance modulus to its
1$\sigma$ lower limit of $\mu = 32.7$, and using a risetime as high
as 20.0d. Even then the model temperature is rather too high. A
similar problem was encountered at both epochs.  Nevertheless, most of
the features of the early-time spectra were quite successfully
reproduced by the model.  Future modelling improvements will include
stratified abundances, plus a better description of the radiation
field.

Only one early-time IR spectrum was obtained.  As with the optical
spectra, this showed that SN~2002er was fairly typical but with some
differences from other events.

An optical spectrum obtained at +215\,d spectrum was modelled using a
simple one-zone nebular code. This provided a fair match to the
observed spectrum.  A $^{56}$Ni mass of 0.69M$_\odot$ was inferred --
consistent with that derived from the light curve
\citep[0.7M$_\odot$][]{gp:04}, and indicating that the fraction of
neutral and triply-ionised iron-group elements is quite small.

The principal contribution of this paper to the study of the SN~Ia
phenomenon has been its provision of an extensive, frequently-sampled
set of early-time spectra.  These spectra span the crucial phase
during which the photosphere recedes from the outermost unburned
layers into the innermost iron-group regions of the core.  This
provides a superb dataset against which explosion scenarios will be
tested via ``tomographic'' spectral synthesis models. Such studies
will be vital in order to understand the significance of the small
differences seen between different type~Ia supernovae.

\section*{Acknowledgments}

We are grateful to all the observers who had to give up
part of their observing time so that SN 2002er could be observed,
and to the staff at the various observatories for efficiently
scheduling/carrying out our service/override requests.
R.K. would additionally like to thank U. Thiele for remeasuring
the arcs used in the CAFOS observations. We also appreciate
A. Pastorello's help in acquiring the AFOSC observations.
Support from the EC Programme `The Physics of Type 
Ia SNe' (HPRN-CT-2002-00303) is acknowledged.
This work is based on observations collected at the European
Southern Observatory, Chile (ESO N$^o$ 69.D-0670), the Italian
Telescopio Nazionale Galileo (TNG), La Palma, and the Isaac 
Newton (INT), and William Herschel (WHT) Telescopes of the
Isaac Newton Group, La Palma, The TNG is operated on the island of La
Palma by the Centro Galileo Galilei of INAF (Instituto Nazionale di
Astrofisica) at the Spanish Observatorio del Roque de los Muchachos of
the Instituto de Astrofisica de Canarias. The INT and WHT are
operated on the island of La Palma by the Isaac Newton Group (ING) in
the Spanish Observatorio del Roque de los Muchachos of the Instituto
de Astrofisica de Canarias. Also based in part on observations 
collected at the CAHA at Calar Alto, operated 
jointly by the Max-Planck Institut f\"{u}r Astronomie and the Instituto 
de Astrofisica de Andalucia (CSIC).


\begin{thebibliography}{99}

\bibitem[\protect\citeauthoryear{Abbott \& Lucy}{1985}]{al85}
Abbott~D.C., Lucy~L.B., 1985, ApJ, 288, 679

\bibitem[\protect\citeauthoryear{Ayani \& Yamaoka}{1998}]{ay:98}
Ayani,K., Yamaoka, H., 1998, IAUC 6878

\bibitem[\protect\citeauthoryear{Benetti et al.}{2004a}]{benetti:04a}
Benetti, S., Meikle, P., Stehle, M. et al., 2004a, MNRAS, 348, 261

\bibitem[\protect\citeauthoryear{Benetti et al.}{2004b}]{benetti:04b}
Benetti, S.et al., 2004b, ApJL (submitted)

\bibitem[\protect\citeauthoryear{Bowers et al.}{1997}]{bowers:97}
Bowers, E.J.C., Meikle, W.P.S., Geballe, T.R., Walton, N.A., Pinto, P.A.,
Dhillon, V.S., Howell, S.B., Harrop-Allin, M.K., 1997, MNRAS, 290, 663

\bibitem[\protect\citeauthoryear{Branch et al.}{1995}]{branch:95}
Branch, D., Livio, M., Yungelson, L.R. et al., 1995, PASP, 107, 1019

\bibitem[\protect\citeauthoryear{Branch et al.}{2003}]{branch:03}
Branch, D., Garnavich, P., Matheson, T. et al.

\bibitem[\protect\citeauthoryear{Cardelli et al.}{1989}]{cardelli:89}
Cardelli, J.A., Clayton, G.C., Mathis, J.S., 1989, ApJ, 345, 245

\bibitem[\protect\citeauthoryear{Drell et al.}{2000}]{drell:00}
Drell, P.S., Loredo, T.J., Wasserman, I., 2000, ApJ, 530, 593 

\bibitem[\protect\citeauthoryear{Riess \etal}{1999}]{ri99} 
Riess~A.G., Filippenko~A.V., Li~W., \etal, 1999, AJ, 118, 2675

\bibitem[\protect\citeauthoryear{Fisher et al.}{1997}]{fisher:97}
Fisher A., Branch D., Nugent P., Baron E., 1997, ApJ, 481, L89

\bibitem[\protect\citeauthoryear{Hamuy et al.}{2002}]{hamuy:02}
Hamuy, M., Maza, J., Pinto, P.A. et al., 2002, ApJ, 124, 417

\bibitem[\protect\citeauthoryear{Hamuy et al.}{2003}]{hamuy:03}
Hamuy, M., Phillips, M. M., et al. 2003, Nature, 424, 651

\bibitem[\protect\citeauthoryear{Hatano \etal}{1999}]{ha99}
Hatano~K., Branch~D., Fisher~A., \etal, 1999, ApJ, 525, 881

\bibitem[\protect\citeauthoryear{Hernandez et al.}{2000}]{hernandez:00}
Hernandez, M., Meikle, W.P.S., Aparicio, A. et al., 2000, MNRAS, 319, 223

\bibitem[\protect\citeauthoryear{Hillebrandt \& Niemeyer}{2000}]{hn:00}
Hillebrandt \& Niemeyer, 2000, ARA\& A, 38, 191

\bibitem[\protect\citeauthoryear{Jha et al.}{1999}]{jha:99}
Jha, S., Garnavich, P., Kirshner, R. et al., ApJS, 125, 73

\bibitem[\protect\citeauthoryear{Khokhlov}{1991}]{khokhlov:91}
Khokhlov, A.M., 1991, A\&A, 245, 114

\bibitem[\protect\citeauthoryear{Kotak et al.}{2004}]{kotak:04}
Kotak, R., Meikle, W.P.S., Adamson, A., Leggett, S.K., 2004, MNRAS, 354, L13

\bibitem[\protect\citeauthoryear{Leibundgut}{2000}]{leibundgut:00}
Leibundgut, B., 2000, A\&AR, 10, 179

\bibitem[\protect\citeauthoryear{Lentz et al.}{2000}]{lentz:00}
Lentz, E.J., Baron, E., Branch, D. et al., 2000, ApJ, 530, 966

\bibitem[\protect\citeauthoryear{Lucy}{1999}]{l99} Lucy~L.B., 1999,
A\&A, 345, 211

\bibitem[\protect\citeauthoryear{Marion et al.}{2003}]{marion:03}
Marion, G.H., H\"{o}flich, P., Vacca, W.D. et al., 2003, ApJ, 591, 316

\bibitem[\protect\citeauthoryear{Mazzali \& Lucy}{1993}]{ml93}
Mazzali~P.A., \& Lucy~L.B., 1993, A\&A, 279, 447

\bibitem[\protect\citeauthoryear{Mazzali et al.}{1993}]{mazzali:93}
Mazzali, P.A. et al., 1993

\bibitem[\protect\citeauthoryear{Mazzali et al.}{1995}]{mazzali:95}
Mazzali, P.A., Danziger, I.J., \& Turatto, M., 1995, A\& A, 297, 509

\bibitem[\protect\citeauthoryear{Mazzali et al.}{1998}]{mazzali:98}
Mazzali, P.A., Cappellaro, E., Danziger, I.J., Turatto, M., Benetti, S.,
1998, ApJ, 499, L49

\bibitem[\protect\citeauthoryear{Mazzali}{2000}]{m00} Mazzali~P.A.,
2000, A\&A, 363, 705

\bibitem[\protect\citeauthoryear{Nomoto et al.}{1984}]{nomoto:84}
Nomoto, K., Thielemann, F.-K, \& Yokoi, K., 1984, ApJ, 286, 644

\bibitem[\protect\citeauthoryear{Nomoto et al.}{1985}]{nomoto:85}
Nomoto, K. et al., 1985

\bibitem[\protect\citeauthoryear{Nugent et al.}{1995}]{nugent:95}
Nugent, P., Phillips, M., Baron, E., et al., 1995, ApJ, 455, L147

\bibitem[\protect\citeauthoryear{Patat et al.}{1996}]{patat:96}
Patat, F., Benetti, S., Cappellaro, E. et al. 1996, MNRAS, 278, 111

\bibitem[\protect\citeauthoryear{Phillips}{1993}]{ph93}
Phillips~M.M., 1993, ApJ, 413, L105

\bibitem[\protect\citeauthoryear{Phillips et al.}{1999}]{phillips:99}
Phillips, M.M., Lira, P., Suntzeff, N.B. et al., 1999, AJ, 118, 1766

\bibitem[\protect\citeauthoryear{Pignata et al.}{2004}]{gp:04}
Pignata, G., et al., 2004, MNRAS (accepted)

\bibitem[\protect\citeauthoryear{Riess et al.}{1998}]{riess:98}
Riess, A. G., et al., 1998, AJ, 116, 1009 

\bibitem[\protect\citeauthoryear{Riess et al.}{1999}]{riess:99}
Riess, A., Filippenko, A.V., Li., W., et al. 1999, ApJ, 118, 2675

\bibitem[\protect\citeauthoryear{R\"opke \& Hillebrandt}{2004}]{roepke:04}
R\"opke \& Hillebrandt, 2004, A\& A (in press) astro-ph/0411667

\bibitem[\protect\citeauthoryear{Ruiz-Lapuente et al.}{2004}]{ruizlapuente:04}
Ruiz-Lapuente, P. et al. astro-ph/0410673 

\bibitem[\protect\citeauthoryear{Salvo et al.}{2001}]{salvo:01}
Salvo, M.E., Cappellaro, E., Mazzali, P.A., et al. 2001, MNRAS, 321, 254

\bibitem[\protect\citeauthoryear{Shortridge}{2002}]{shortridge:02}
Shortridge K., 2002, Star link User Note 86. 20

\bibitem[\protect\citeauthoryear{Smartt et al.}{2004}]{ss:02}
Smartt, S.J., Patat, F., Meikle, P., \& Araujo, S., 2002, IAUC 7961

\bibitem[\protect\citeauthoryear{Stehle et al.}{2004}]{stehle:04}
Stehle, M., Mazzali, P.A., Benetti, S., Hillebrandt, W., 2004,
MNRAS (submitted) astro-ph/0409342

\bibitem[\protect\citeauthoryear{Nomoto \etal}{1984}]{n84} Nomoto~K.,
Thielemann F.-K., Yokoi K., 1984, ApJ, 286, 644

\bibitem[\protect\citeauthoryear{Wheeler et al.}{1998}]{wheeler:98}
Wheeler, J.C., H\"{o}flich, P., Harkness, R.P., Spyromilio, J., 1998, ApJ, 496, 908


\bibitem[\protect\citeauthoryear{Wood-Vasey et al.}{2002}]{wv:02}
Wood-Vasey, W.M., Li, W.D., Swift, B., \& Ganeshalingam, M., 2002, IAUC 7959


\end{thebibliography}
\end{document}